\documentclass[11pt]{article}
\usepackage{graphicx}
\usepackage{moriond,epsfig}

\newcommand{\BABARPubYear}    {01}

\newcommand{\BABARProcNumber} {24}
\newcommand{\SLACPubNumber} {8833}

\def\babar{\mbox{\slshape B\kern-0.1em{\small A}\kern-0.1em
    B\kern-0.1em{\small A\kern-0.2em R}}}
\def\pep2{PEP-II}
\def\Y#1S{\ensuremath{\Upsilon{(#1S)}}}
\def\FourS {\Y4S}

\def\epem       {\ensuremath{e^+e^-}}
\def\CP                 {\ensuremath{C\!P}}
\def\BF{$B$ Factory}
\def\mes        {\mbox{$m_{\rm ES}$}}
\newcommand{\etaCP}{\ensuremath{\eta_{\CP}}}

\def\piz   {\ensuremath{\pi^0}}

\def\pip   {\ensuremath{\pi^+}}
\def\pim   {\ensuremath{\pi^-}}

\def\Kbar  {\kern 0.2em\overline{\kern -0.2em K}{}}

\def\Km    {\ensuremath{K^-}}
\def\KS    {\ensuremath{K^0_{\scriptscriptstyle S}}} 
\def\KL    {\ensuremath{K^0_{\scriptscriptstyle L}}} 
\def\Kstarz  {\ensuremath{K^{*0}}}

\def\Kzb   {\ensuremath{\Kbar^0}}
\def\KzKzb {\ensuremath{K^0 \kern -0.16em \Kzb}}

\def\D     {\ensuremath{D^+}}
\def\Dbar  {\kern 0.2em\overline{\kern -0.2em D}{}}

\def\Dzb   {\ensuremath{\Dbar^0}}
\def\DzDzb {\ensuremath{D^0 {\kern -0.16em \Dzb}}}
\def\Dstar   {\ensuremath{D^*}}

\def\Bz    {\ensuremath{B^0}}
\def\B     {\ensuremath{B}}
\def\Bbar  {\kern 0.18em\overline{\kern -0.18em B}{}}
\def\Bb    {\ensuremath{\Bbar}}
\def\Bzb   {\ensuremath{\Bbar^0}}
\def\Bu    {\ensuremath{B^+}}
\def\Bub   {\ensuremath{B^-}}

\def\BzBzb {\ensuremath{B^0 {\kern -0.16em \Bzb}}}

\def\jpsi  {\ensuremath{{J\mskip -3mu/\mskip -2mu\psi\mskip 2mu}}} 
\def\psitwos {\ensuremath{\psi{(2S)}}}
\mathchardef\Upsilon="7107
\def\Y#1S{\ensuremath{\Upsilon{(#1S)}}}

\def\FourS {\Y4S}

\mathchardef\Deltares="7101
\mathchardef\Xi="7104
\mathchardef\Lambda="7103
\mathchardef\Sigma="7106
\mathchardef\Omega="710A
\def\Deltabar   {\kern 0.25em\overline{\kern -0.25em \Deltares}{}}
\def\Lbar {\kern 0.2em\overline{\kern -0.2em\Lambda\kern 0.05em}\kern-0.05em{}}
\def\Sigbar{\kern 0.2em\overline{\kern -0.2em \Sigma}{}}
\def\Xibar{\kern 0.2em\overline{\kern -0.2em \Xi}{}}
\def\Obar{\kern 0.2em\overline{\kern -0.2em \Omega}{}}
\def\Nbar{\kern 0.2em\overline{\kern -0.2em N}{}}
\def\Xbar{\kern 0.2em\overline{\kern -0.2em X}{}}

\def\bflav{\ensuremath{B_{\rm flav}}}

\def\invfb   {\ensuremath{\mbox{\,fb}^{-1}}}

\newcommand{\epjc}      [1]  {{Eur.\ Phys.\ Jour.\ C~{\bf #1}}}

\newcommand{\nim}       [1]  {{Nucl.\ Instr.\ and Methods~{\bf #1}}}

\def\ev   {\ensuremath{\rm \,e\kern -0.08em V}}
\def\kev  {\ensuremath{\rm \,ke\kern -0.08em V}} 
\def\mev  {\ensuremath{\rm \,Me\kern -0.08em V}} 
\def\gev  {\ensuremath{\rm \,Ge\kern -0.08em V}} 
\def\gevc {\ensuremath{{\rm \,Ge\kern -0.08em V\!/}c}} 
\def\tev  {\ensuremath{\rm \,Te\kern -0.08em V}}
\def\mevc {\ensuremath{{\rm \,Me\kern -0.08em V\!/}c}} 
\def\gevcc{\ensuremath{{\rm \,Ge\kern -0.08em V\!/}c^2}} 
\def\mevcc{\ensuremath{{\rm \,Me\kern -0.08em V\!/}c^2}}

\def\cm   {\ensuremath{\rm \,cm}}

\def\mum  {\ensuremath{\,\mu\rm m}} 

\def\invfb   {\ensuremath{\mbox{\,fb}^{-1}}}
\def\mus  {\ensuremath{\rm \,\mus}}

\def\ps   {\ensuremath{\rm \,ps}}

\def\stwob{\ensuremath{\sin\! 2 \beta   }}

\def\mistag{\ensuremath{w}}

\def\deltaz{\ensuremath{{\rm \Delta}z}}
\def\deltat{\ensuremath{{\rm \Delta}t}}
\def\deltamd{\ensuremath{{\rm \Delta}m_d}}

\def\result {\ensuremath{\stwob=0.34\pm 0.20\, {\rm (stat)} \pm 0.05\, {\rm (syst)}}}

\long\def\inst#1{\par\nobreak\kern 4pt\nobreak
    {\it #1}\par\vskip 10pt plus 3pt minus 3pt}

\begin{document}
{\pagestyle{empty}

\begin{flushright}
SLAC-PUB-\SLACPubNumber \\
\babar-PROC-\BABARPubYear/\BABARProcNumber \\
May, 2001 \\
\end{flushright}

\par\vskip 4cm

\begin{center}
\Large \bf Measurements of CP violation, mixing and lifetimes in \B\ meson decays with the \babar\ experiment at \pep2.
\end{center}
\bigskip

\begin{center}
\large 
Riccardo Faccini\\
University of California San Diego and Universit\`a ``La Sapienza'' Roma \\
(for the \babar\ Collaboration)
\end{center}
\bigskip \bigskip

\begin{center}
\large \bf Abstract
\end{center}
The \babar\ detector, which operates at the SLAC \pep2\ asymmetric \epem\ 
collider at energies near the \FourS\ resonance has collected about 23M \B\Bb\ pairs in year 2000. Based on this data sample, 
we present a first study of \stwob, with samples 
of $\Bz \to \jpsi \KS$ , $\Bz \to \psitwos \KS$  and $\Bz \to \jpsi \KL$ decays.
The measured value is \result. 
In addition, we present preliminary measurements of charged and neutral \B\ meson lifetimes and $\Bz\Bzb$  
oscillation frequency.

\vfill
\begin{center}
Contributed to the Proceedings of the \\
36th Rencontres de Moriond on Weak Interactions \\
10-17 Mar 2001, Les Arcs, France
\end{center}

\vspace{1.0cm}
\begin{center}
{\em Stanford Linear Accelerator Center, Stanford University, 
Stanford, CA 94309} \\ \vspace{0.1cm}\hrule\vspace{0.1cm}
Work supported in part by Department of Energy contract DE-AC03-76SF00515.
\end{center}

\newpage
\section{Introduction}
The primary goal of the \babar\ experiment at \pep2\ is to over-constrain the Unitarity Triangle.
The sides of this triangle can be measured through non-\CP\ violating physics, such as  
$V_{ub}$, $V_{cb}$, $V_{td}$ measurements, 
while its angles are accessible through \CP\ violating processes~\cite{PhysBook}.
\section{PEP-II}
The \pep2\ $B$ Factory~\cite{BabarPub0018} is an \epem\ colliding beam storage ring complex on the SLAC site 
designed to produce a luminosity of at least 3x$10^{33} \cm^{-2}s^{-1}$ at 
a center--of--mass energy of 10.58\gev, the mass of the \FourS\ resonance. In the 2000 run, 
the achieved average luminosity was 3.3x$10^{33} \cm^{-2}s^{-1}$. 
The total collected luminosity was about $23 \invfb$.
The machine is asymmetric with a High Energy Ring (HER) for the 9.0\gev\ electron beam
and a Low Energy Ring (LER) for the 3.1\gev\ positron beam. This corresponds to
$\rm {\beta\gamma}$=0.56 and makes it possible to measure time dependent \CP\ violating asymmetries. 
It corresponds to an average separation of $\rm {\beta\gamma c \tau}$=250\mum\ 
between the two $B$ mesons vertices.

\section{\mbox{\sl B\hspace{-0.4em} {\small\sl A}\hspace{-0.4em} \sl B\hspace{-0.4em} {\small\sl A\hspace{-0.1em}R}}}

\title{Measurements of CP violation, mixing and lifetimes in \B\ meson decays with the \babar\ experiment at \pep2.}

\subsection{Detector description}
The \babar\ detector is described in \cite{NIM}.
The volume within the \babar\ superconducting solenoid, which produces a 1.5 T axial magnetic
field, consists of: a five layer silicon strip vertex detector (SVT), a central drift
chamber (DCH), a quartz-bar Cherenkov radiation detector (DIRC) and a CsI
crystal electromagnetic calorimeter (EMC). Two layers of
cylindrical resistive plate counters (RPCs) are located between the barrel
calorimeter and the magnet cryostat. All the detectors located inside the
magnet have full acceptance in azimuth. The integrated flux return (IFR) outside the cryostat is
composed of 18 layers of steel, which successively increase in thickness away from the
interaction point, and are instrumented with 19 layers of planar RPCs in the barrel and 18 in the endcaps.
\subsection{Event reconstruction}
Charged particles are detected and their momentum is measured by a combination of 
the DCH and SVT. The  charged particle momentum resolution is approximately given by 
$\left( \delta p_T / p_T \right)^2 =  (0.0015\, p_T)^2 + (0.005)^2$, where 
$p_T$ is in \gevc.  The SVT, with a typical resolution of 
10\mum\ per hit, provides excellent vertex resolution  both in 
the transverse plane and in $z$.  The vertex resolution in $z$ is typically 50\mum\ for 
a fully reconstructed \B\ meson and of order 100\mum\ for the distance 
among the two \B\ mesons when only one is fully reconstructed.
Leptons and hadrons are identified using a combination of measurements
from all the \babar\ components, including 
the energy loss ${\rm d}E/{\rm d}x$ in the helium-based 
gas of the DCH (40 samples maximum) and in the silicon of the SVT (5 samples
maximum). 
Electrons and photons are identified in the barrel and the forward regions 
by the EMC, and muons are identified in the IFR. In the barrel region the DIRC
provides excellent kaon identification over the full 
momentum range above 250\mev/c.
\section{${\sin\! 2 \beta   }$ measurement}
In \epem\ storage rings operating at  the \FourS\ resonance a \BzBzb\ pair
produced in a \FourS\ decay 
evolves in a coherent $P$-wave until one of the \B\ mesons decays. 
If one of the \B\ mesons ($B_{tag}$) can be 
ascertained to decay to a state of known flavor at a certain time $t_{tag}$, 
the other \B\  ($B_{CP}$) is {\it at that time} known to be of the opposite flavor.
For the measurement of \stwob, $B_{CP}$ is fully reconstructed in a \CP\ 
eigenstate ($\jpsi \KS$, $\psitwos \KS$ or $\jpsi \KL$).
By measuring the proper time interval $\deltat = t_{CP} - t_{tag}$ from the $B_{tag}$ decay time 
to the decay of the $B_{CP}$ ($t_{CP}$), it is possible to determine the time evolution 
of the initially pure \Bz\ or \Bzb\ state:
\begin{equation}
\label{eq:TimeDep}
        f_\pm(\, \deltat \, ; \,  \Gamma, \, \deltamd, \, {\cal {D}} \sin{ 
2 \beta } )  = {\frac{1}{4}}\, \Gamma \, {\rm e}^{ - \Gamma \left| \deltat 
\right| }\, \left[  \, 1 \, \mp \, {\cal {D}}\etaCP \sin{ 2 \beta } \times \sin{ \deltamd \, \deltat } \,  \right]\ ,
\end{equation}
where the $+$ or $-$  sign 
indicates whether the 
$B_{tag}$ is tagged as a \Bz\ or a \Bzb, respectively.  The dilution factor ${\cal {D}}$ is given by 
$ {\cal {D} } = 1 - 2 \mistag$, where $\mistag$ is the mistag fraction, {\it i.e.}, the 
probability that the flavor of the tagging \B\ is identified incorrectly. 
\etaCP\ is the \CP\ eigenstate of the final state and it is $\etaCP=-1$ for the $\jpsi \KS$ and  $\psitwos \KS$ modes,
$\etaCP=+1$ for the $\jpsi \KL$ mode. Although less pure, the $\jpsi \KL$ mode is very important because the oscillation is expected to be opposite to the other ones. 
A direct \CP\ violation term proportional to $\cos {  \deltamd \, \deltat }$ 
could arise from the interference between two decay mechanisms with different
weak phases.  In the Standard Model, we consider that the dominant diagrams for the decay  modes 
have no relative weak phase, so no such term is expected.
\par
To account for the finite resolution of the detector,
the time-dependent distributions $f_\pm$ for \Bz\ and \Bzb\ tagged events 
(Eq.~\ref{eq:TimeDep}) must be convoluted with 
a time resolution function ${\cal {R}}( \deltat ; \hat {a} )$:
\begin{equation}
\label{eq:Convol}
        {\cal F}_\pm(\, \deltat \, ; \, \Gamma, \, \deltamd, \, {\cal {D}}\etaCP \sin{ 2 \beta }, \hat {a} \, )  = 
f_\pm( \, \deltat \, ; \, \Gamma, \, \deltamd, \, {\cal {D}}\etaCP \sin{ 2 \beta } \, ) \otimes 
{\cal {R}}( \, \deltat \, ; \, \hat {a} \, ) \ ,
\end{equation}
where $\hat {a}$ represents the set of parameters that describe the resolution function.  
\par

Finally, the time-dependent distributions need to account 
for the background with additional parameters that characterize 
both its sample composition and the \deltat\ distribution.
The $\etaCP=-1$ background components are parametrized with an ARGUS function and a small peaking component in the energy substituted mass \mes. While the ARGUS component is extracted from data, the peaking one is derived from data.
The $\etaCP=+1$ background is made of a component from \B\ decays in \jpsi, whose parameters are taken from MC, and the other sources which are characterized using the \jpsi\ mass sidebands.

Since no time-integrated \CP\ asymmetry effect is expected, an
analysis of the time-dependent asymmetry is necessary.  


\subsection{Analysis}
For this analysis, published in \cite{PRL}, we use a sample of $23 \invfb$ of data recorded
in year 2000, 
of which $2.6 \invfb$ was recorded 40\mev\ below the \FourS\ resonance (off-resonance data). 
\par
The measurement of the \CP-violating asymmetry has five main components~:
\begin{itemize}
\item
Selection of the signal $\Bz/\Bzb \to \jpsi \KS$,
 $\Bz/\Bzb \to \psitwos \KS$ and $\Bz/\Bzb \to \jpsi \KL$ events, as described in detail
in~\cite{BabarPub0005}.   
\par
The selection of the two modes with the \KS\ can profit from the fact that there are two discriminating kinematic
variables, ${\rm \Delta} E$, the difference between the reconstructed and
expected \B\ meson energy measured in the center--of--mass frame, and \mes,
the beam--energy substituted mass (see 
Fig.~\ref{fig:jks00}). The background is mainly coming from combinatorics in continuum and other \B\ decays and its properties can be estimated from the \mes\ sidebands.

The \jpsi\KL sample is instead  less pure because the the \KL\ momentum is not reconstructed. The \B\ mass constraint is therefore imposed and only one discriminating variable is left ( ${\rm \Delta} E$ ) as shown in Fig.~\ref{fig:jks00}.
The background is mainly due to other $\B\to\jpsi X$ decays and its properties are determined on MC.

Signal event yields and purities, determined 
from a fit to the \mes\ distributions after selection 
on ${\rm \Delta} E$, are summarized in
Table~\ref{tab:result}. 
\begin{figure}[t]
\begin{center}
 \mbox{\epsfig{file=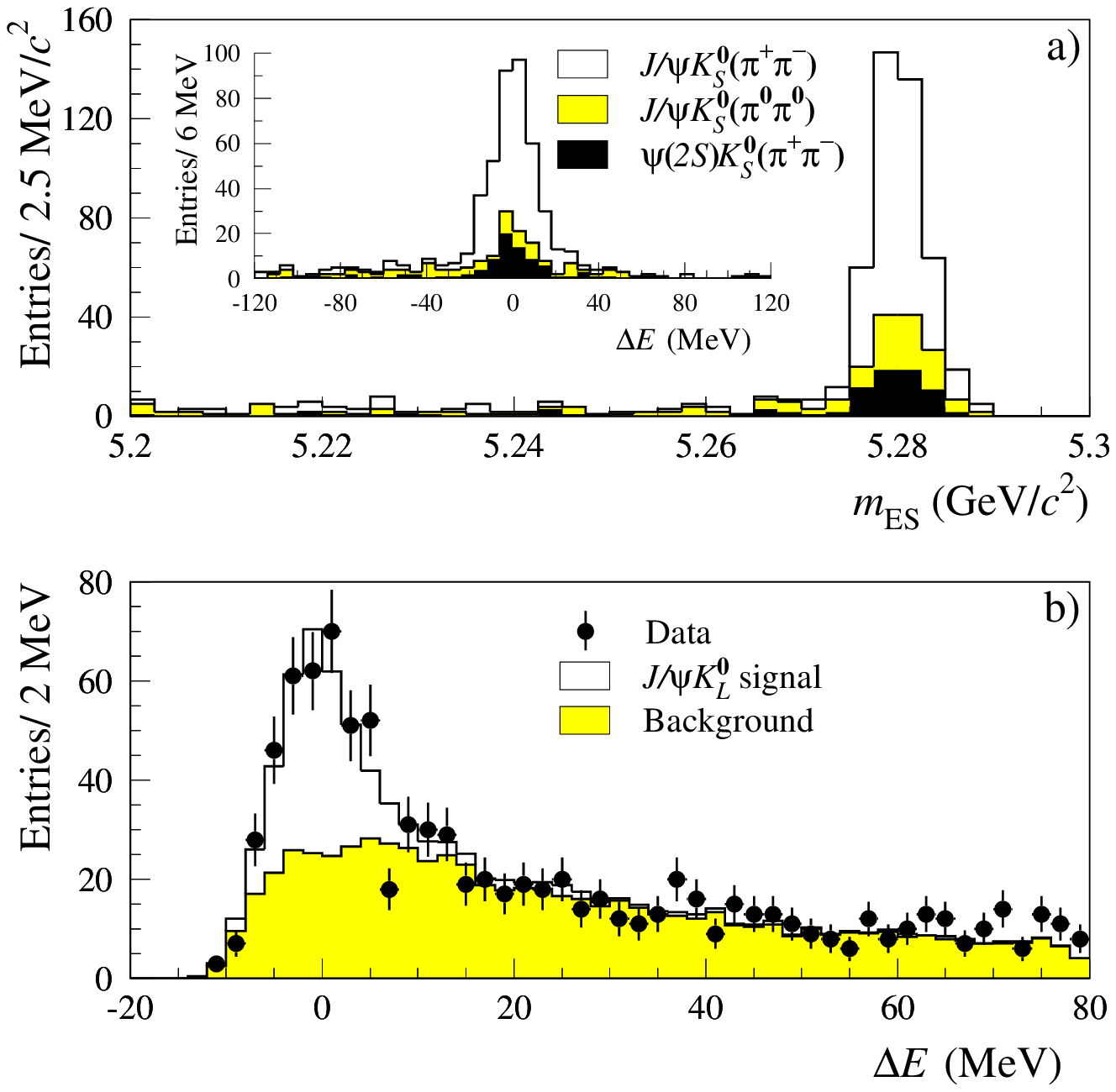,height=8cm}} 
 \caption{\it $\jpsi \KS$ ($\KS \to \pi^+ \pi^-, \pi^0 \pi^0$)  and
$\psi \KS$ ($\KS \to \pi^0 \pi^0$) signal (top) and $\jpsi \KL$ signal (bottom) .
\label{fig:jks00}}
\end{center}
\end{figure}

\begin{table}[!htb]
\begin{center} 
\caption{ 
Number of tagged events, signal purity and result of fitting for \CP\ asymmetries in 
the full \CP\ sample and in 
various subsamples, as well as in the $B_{\rm flav}$ and charged $B$ control samples.  
}
\label{tab:result} 
\vspace{0.2cm}
\begin{tabular}{|c|c|c|c|} \hline
 Sample                                  & $N_{\rm tag}$    & Purity (\%)    &  \stwob  \\ \hline \hline
$\jpsi\KS$, $\psitwos\KS$                & $273$        & $96\pm1$       &  0.25$\pm$0.22   \\ 
$\jpsi \KL$                              & $256$        & $39\pm6$       &  0.87$\pm$0.51   \\ 
\hline
 Full \CP\ sample                        & $529$        & $69\pm2$       &  0.34$\pm$0.20   \\ 
\hline
\hline
$\jpsi\KS$, $\psitwos\KS$ only & & & \\
\hline
$\ \jpsi \KS$ ($\KS \to \pi^+ \pi^-$)    & $188$        & $98\pm1$       &  0.25$\pm$0.26   \\ 
$\ \jpsi \KS$ ($\KS \to \pi^0 \pi^0$)    & $41$         & $85\pm6$       &  -0.05$\pm$0.66  \\ 
$\ \psi(2S) \KS$ ($\KS \to \pi^+ \pi^-$) & $44$         & $97\pm3$       &  0.40$\pm$0.50   \\ 
\hline 
$\ $ {\tt Lepton} tags                   & $34$         &  $99\pm2$      &  0.07$\pm$0.43   \\ 
$\ $ {\tt Kaon} tags                     & $156$        &  $96\pm2$      &  0.40$\pm$0.29    \\ 
$\ $ {\tt NT1} tags                      & $28$         &  $97\pm3$      &  -0.03$\pm$0.67    \\ 
$\ $ {\tt NT2} tags                      & $55$         &  $96\pm3$      &  0.09$\pm$0.76    \\ 
\hline\hline
$B_{\rm flav}$ sample                    & $4637$       & $86\pm1$       &  0.03$\pm$0.05     \\
\hline 
Charged $B$ sample                       & $5165$       & $90\pm1$       &  0.02$\pm$0.05     \\ \hline

\end{tabular} 
\end{center}
\end{table}

\item
Selection of decays in flavor eigenstate (\bflav).
\Bz\ candidates are formed by combining a \Dstar\ or \D\ 
with a \pip, $\rho^+$ $(\rho^+\to\pip\piz)$, $a_1^+$ $(a_1^+\to\pip\pim\pip)$,
or by combining a \jpsi\ candidate
with a \Kstarz\ $(\Kstarz\to\Km\pip)$\cite{BabarPub0008}. 
Their background is mainly due to conbinatorics and can be studied in the \mes\ sidebands. 
Yields and purities are also summarized in Table~\ref{tab:result}.
\item
Measurement of the distance \deltaz\ between the vertex of the reconstructed $B$ meson ($B_{\rm rec}$) 
and the vertex of the flavor-tagging $B$ meson ($B_{\rm tag}$). 

In the reconstruction of the $B_{\rm rec}$ vertex, 
we use all charged daughter tracks.
The vertex for the $B_{\rm tag}$ decay is constructed from all the remaining
tracks in the event. 

In order to reduce bias and tails due to long-lived particles,
\KS\ and $\Lambda^0$ candidates are used as 
input to the fit in place
of their daughters. In addition, tracks consistent with photon conversions 
($\gamma \to e^+ e^-$) are excluded. 
To reduce contributions from charm decay products, 
which bias the determination of the vertex position,
the track with the largest vertex $\chi^2$ contribution greater than 
6 is removed and the fit is redone until no track fails the $\chi^2$ 
requirement or only one track remains.

From the measurement of \deltaz\ and of the \B\ momentum, \deltat\ can be computed.
At an asymmetric-energy \BF, in fact, the proper decay-time difference $\deltat$ is, 
to an excellent approximation, proportional to 
the distance \deltaz\ between  the two \Bz-decay vertices 
along the axis of the boost, 
$\deltat \approx \deltaz / {\rm c} \left< \beta \gamma \right>  $.  
\par
The time resolution function in equation \ref{eq:Convol} is described by a
sum of three Gaussian distributions 
(called the core, tail and outlier components)
with different means and widths:
\begin{eqnarray}
{\cal {R}}( \delta_{\rm t} ; \hat {a} ) &=&  \sum_{k=1}^{2} 
{ \frac{f_k}{\sigma_k\sqrt{2\pi}} \, {\rm exp} 
\left(  - {( \delta_{\rm t}-\delta_{k} )^2 \over 
 2{\sigma_k}^2 }  \right) } \, \,  + 
{ \frac{f_3}{\sigma_3\sqrt{2\pi}} \, {\rm exp} 
\left(  - { \delta_{\rm t}^2 \over 
 2{\sigma_3}^2 }  \right) } \, \, .
\label{eq:vtxresolfunct}
\end{eqnarray}
For the core and tail Gaussians, 
the widths are scaled by the event-by-event measurement
error $\sigma_{\deltat}$ derived from the vertex fits:
$\sigma_{1,2}={\cal S}_{1,2}\times\sigma_{\deltat}$.
In data, approximately 65\% of the area of the resolution function is
in the core Gaussian.
The width of the core Gaussian is approximately 
110\mum\ or 0.7\ps;
the width of the tail Gaussian is approximately 
300\mum\ or 1.8\ps. 
The third Gaussian has a fixed width of 8~ps and no offset; it  accounts
for the fewer than 1\% of events with incorrectly reconstructed 
vertices.

\item
Determination of the flavor of the $B_{tag}$.
\par
Each event with a \CP\ candidate is assigned a $\Bz$ or $\Bzb$ tag if 
the rest of the event ({\it i.e.,} with the daughter
tracks of the $B_{CP}$ removed) satisfies the criteria from one of several 
tagging categories. 
In other words, a \Bz\ tag indicates that the $B_{CP}$ candidate was in a \Bzb\ state 
at $\deltat=0$; a \Bzb\ tag indicates that the $B_{CP}$ candidate was in a \Bz\ state.
\par
Two tagging categories rely on the presence of a fast lepton 
({\tt Lepton} category) and/or one or more charged kaons in the event
({\tt Kaon} category).  
Two categories, called neural network categories ({\tt NT1} and {\tt NT2}), are based upon
the output value of a neural network algorithm applied to events 
that have not already been assigned to lepton or kaon tagging categories.

The figure of merit for each tagging category is the effective tagging efficiency  
$Q_i = \varepsilon_i \, \left( 1 - 2\mistag_i \right)^2$, where $\varepsilon_i$ 
is the fraction of events assigned to category $i$ and 
$\mistag_i$ is the mistag fraction. The effective tagging efficiency as evaluated in data is summarized in 
Tab.~\ref{tab:TagMix:mistag}.

\begin{table}[!tb]
\begin{center}
\caption{
Mistag fractions measured from 
a maximum-likelihood fit to the time distribution for the fully-reconstructed \Bz\ sample.
The uncertainties on $\varepsilon$ and $Q$ are statistical only.
} 
\vspace{0.2cm}
\begin{tabular}{|c|c|c|c|c|}  \hline 
Category     & $\varepsilon$ (\%) & $\mistag$ (\%) & $\Delta\mistag$ (\%) & $Q$ (\%)       \\ \hline\hline 
{\tt Lepton} & $10.9\pm 0.4$ & $11.6\pm2.0$ & $3.1\pm 3.1$  &  $ 6.4\pm 0.7$  \\ 
{\tt Kaon}   & $36.5\pm 0.7$ & $17.1\pm1.3$ & $-1.9\pm 1.9$ &  $15.8\pm 1.3$  \\ 
{\tt NT1}    &  $7.7\pm 0.4$ & $21.2\pm2.9$ & $7.8\pm 4.2$  &  $ 2.6\pm 0.5$  \\ 
{\tt NT2}    & $13.7\pm 0.5$ & $31.7\pm2.6$ & $-4.7\pm 3.5$ &  $ 1.8\pm 0.5$  \\  \hline \hline
All          & $68.9\pm 1.0$ &              &               &  $26.7\pm 1.6$  \\  \hline 
\end{tabular} 
\end{center}
\label{tab:TagMix:mistag}
\end{table}

\item 
The mistag fractions and the tagging efficiencies 
obtained by combining the results from maximum likelihood fits to the
time distributions in the \Bz\
hadronic and semileptonic samples are summarized 
in Table~\ref{tab:TagMix:mistag}. 
\item
Extraction of the amplitude of the \CP\ asymmetry and the value of \stwob\
with an unbinned maximum likelihood fit.  In order to extract as much information from the data itself and properly account for correlation,
the fit is performed simultaneously to the \CP\ and the flavor eigenstates. There are 35 parameters free in the fit:
\begin{itemize}
\item
{\bf Value of {\boldmath \stwob};}
\item
{\bf Signal resolution function:}
Nine parameters $\hat a_i$ to describe the resolution function for the signal, being scale factors $S_{1,2}$ for the 
event-by-event \deltaz\ resolution errors of the core and tail Gaussian
components, individual core biases $\delta_{1,i}$ per tagging category and a common tail bias $\delta_2$, and the tail $f_1$ 
and outlier $f_3$
fractions; the width of the outlier component is taken to be a fixed 8\ps\ with zero bias;
\item
{\bf Signal dilutions:}
Eight parameters to describe the measured average dilutions $\langle {\cal D}_i\rangle$ and dilution
differences $\delta {\cal D}_i$ in each tagging category.
\item
{\bf Background resolution function:}
Three parameters are used to describe a common resolution function for all non-peaking backgrounds, 
which is taken as a single Gaussian distribution with a scale factor $S_1$ for the event-by-event 
\deltaz\ errors and an common bias $\delta_1$, and an outlier fraction $f_3$; the width of the 
outlier component is taken to be a fixed 8\ps\ with zero bias;
\item
{\bf $B_{\rm flav}$ background properties:} 
A total of 13 parameters describe the $B_{\rm flav}$ background properties.
We make several assumptions to simplify the parameterization of
the background contributions  and assign a corresponding systematic
uncertainty.  The mixing background contribution is assumed to be absent, $f^{\rm flav}_{i,3}=0$.
The size of the peaking background is determined from Monte Carlo simulation 
to be $\delta_{\rm peak}^{\rm flav}=1.5 \pm 0.5 \%$ of the signal contribution in each tagging category.  This
contribution is dominantely from $\Bu$ events, so $\deltamd=0$,
$\Gamma_{i,{\rm peak}}^{\rm flav}=\Gamma_{\Bu}$ and $D^{\rm flav}_{i,{\rm peak}}$
are taken from the $\Bu$ data sample.  
The effective dilutions for the prompt ($D^{\rm flav}_{i,1}$, 4 parameters) 
and lifetime ($D^{\rm flav}_{i,2}$, 4 parameters) contributions are allowed to vary.
The relative amount of these two contributions is allowed to vary, independently 
in each tagging category (4 parameters).
For the lifetime contribution, $\Gamma_{i,2}^{\rm flav}$ is assumed to be same for all
tagging categories, giving one free parameter.
 
\item
{\bf $\CP$ background properties:}
One parameter, the fraction of prompt relative to lifetime background, assumed
to be the same for each tagging category, is allowed to float to describe
the $\CP$ background properties.  The effective dilutions of the lifetime
and peaking contribution are set to zero ($D^{\CP}_{i,2}=D^{\CP}_{i,{\rm peak}}=0$),
corresponding to no \CP-asymmetry in the background.  The size and parameters of the peaking
background is again determined from Monte Carlo simulation.  
The fraction of peaking background is $\delta_{\rm peak}^{\CP}=1 \pm 1\%$ of the signal contribution, independent
of tagging category.  This contribution is assumed to have
dilutions and lifetime parameters in common with the signal contribution.
Finally, the lifetime of the lifetime background is assumed to be $\tau_{\Bz}$
in all tagging categories.
\end{itemize}

\begin{figure}[t]
\begin{center}
\begin{tabular}{lr}   
 \mbox{\epsfig{file=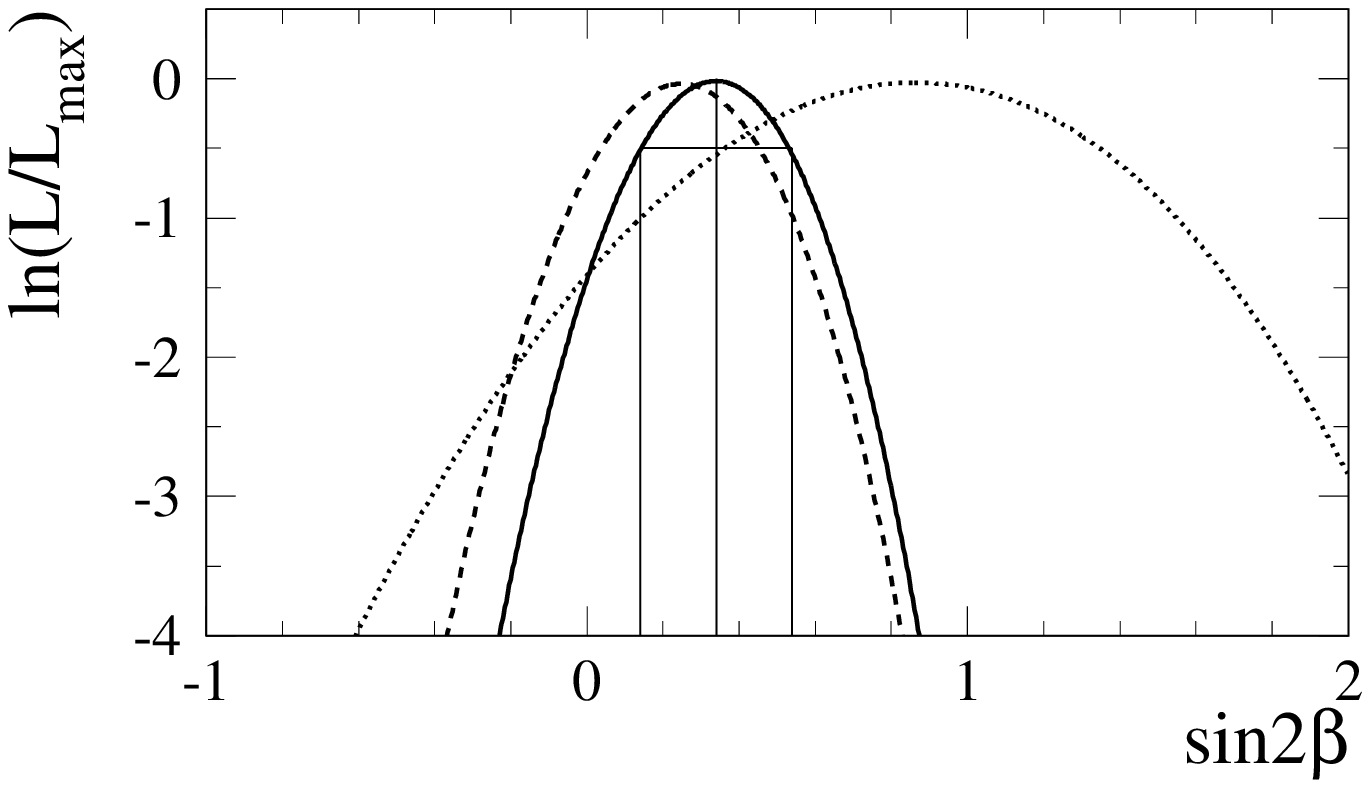, width=8.5cm}}&
 \mbox{\epsfig{file=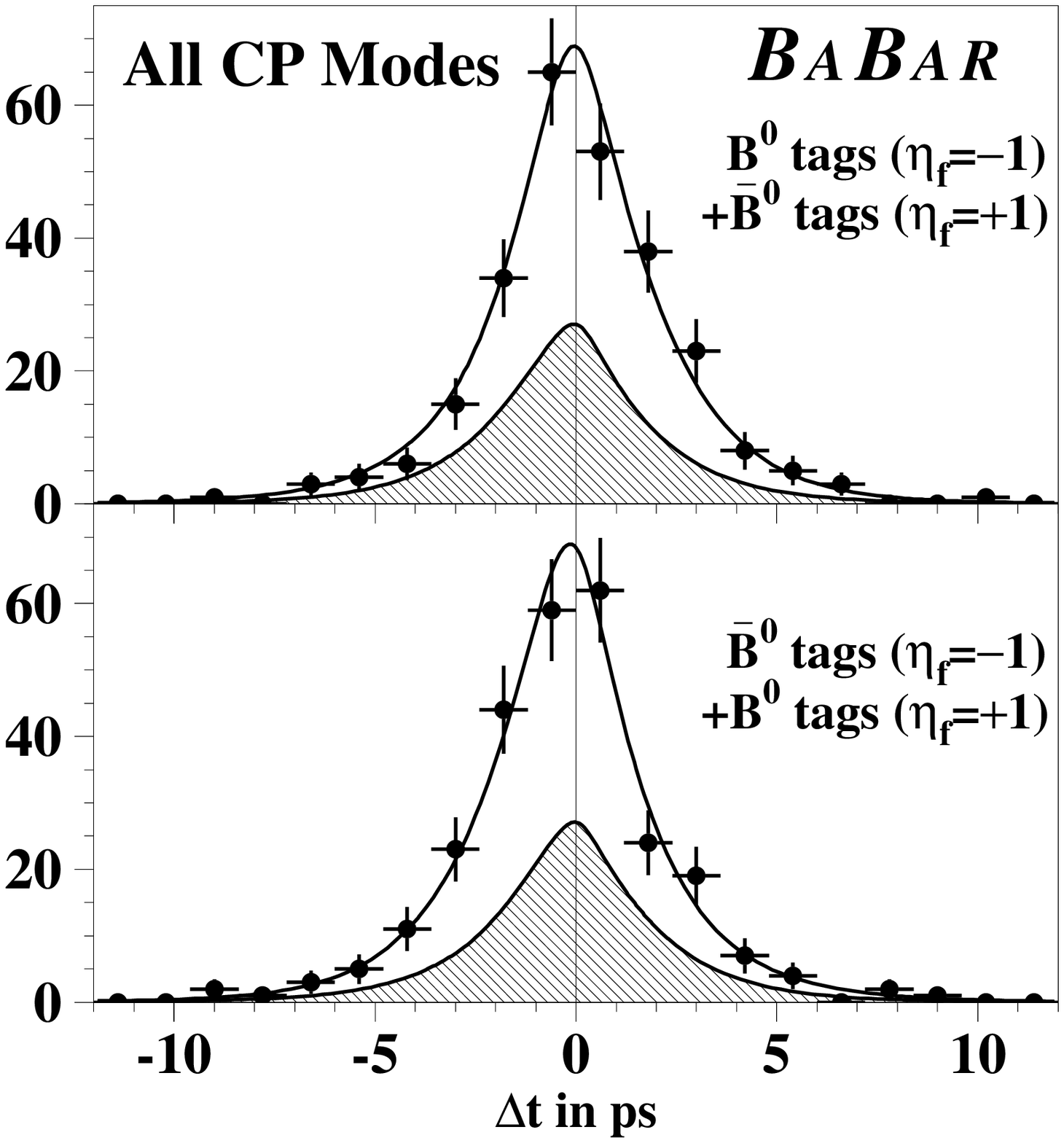, width=7cm}}\\
\end{tabular}
 \caption{\it
      Variation of the log likelihood as a function of \stwob\ (left), for the whole sample (full curve), the 
$\etaCP=-1$ sample (dashed curve) and the $\etaCP=1$ sample (dotted line). 
Distribution of \deltat\ for (a) the \Bz\ tagged events and (b) the  \Bzb\ tagged events 
in the \CP\ sample (right).      \label{fig:likelihood} }
\end{center}
\end{figure}

The maximum-likelihood fit for \stwob, using the full tagged sample,  gives:
\begin{equation}
\result  .
\end{equation}

For this result, the \Bz\ lifetime and \deltamd\ are fixed to the current best 
values~\cite{PDG2000}.
The log likelihood is shown as a function of \stwob\ 
and the \deltat\ distributions for \Bz\ and \Bzb\ tags are shown 
in Fig.~\ref{fig:likelihood}.

\end{itemize}
The dominant sources of systematic error are the assumed parameterization of 
the \deltat\ resolution function (0.04),
due in part to residual uncertainties in the SVT alignment,  
and uncertainties in the level, composition, and \CP\ asymmetry of the background
in the selected \CP\ events (0.02). 
The systematic errors from
uncertainties in $\Delta m_{\Bz}$ and $\tau_{\Bz}$ and from the
parameterization of the background
in the selected $B_{\rm flav}$ sample are found to be negligible. 
An increase of $0.02\,\hbar\ps^{-1}$ in the assumed value for $\Delta m_{\Bz}$ 
decreases \stwob\ by 0.012.
\par
The large sample of reconstructed events allows a number of consistency
checks, including separation of the data by decay mode, tagging
category and $B_{\rm tag}$ flavor. The results of fits to these subsamples
are shown in Table~\ref{tab:result} for the high-purity \KS\ events. Table~\ref{tab:result} also
shows results of fits with the samples of non-\CP\ decay modes, where no statistically
significant asymmetry is found.

\section{Measurements of charged and neutral \B\ meson lifetimes and $\Bz\Bzb$ oscillations}
\label{sec:mixing}
These measurements can be used to test theoretical models of heavy--quark decays 
and to constrain the Unitarity Triangle (via the sensitivity to the value of the CKM
matrix element $V_{td}$).
\par
One \B~($\B_{rec}$) is fully reconstructed in an all-hadronic
($\Bz \to D^{(*)-} \pi^+$, $D^{(*)-} \rho^+$,
$D^{(*)-} a_1^+$, $\jpsi \Kstarz$ and
$\Bu \to \overline{D}^{(*)0} \pip$,
$\jpsi K^+$, $\psitwos K^+$) modes.  The number of selected events and purities are summarized in
Table~\ref{tab:result}. 
\par
The measurement of \deltat\ and, when needed, the tagging of the recoiling \B\ is done in the same way as for the \stwob\ measurement, so that these measurements constitute also a valuable  validation of the \CP\ measurement.
\subsection{Lifetime Measurements}
The \Bz\ and \Bu\ lifetimes are extracted from a simultaneous
unbinned maximum likelihood fit to the $\Delta t$ distributions of the
signal candidates, assuming a common resolution function.
An empirical
description of the $\Delta t$ background shape is assumed, using \mes\
sidebands with independent
parameters for neutral and charged mesons. Fig. \ref{fig:lifetime}
shows the $\Delta t$ distributions with the fit result superimposed.
\begin{figure}[htb]
\begin{center}
\begin{tabular}{lr}   
  \epsfig{file=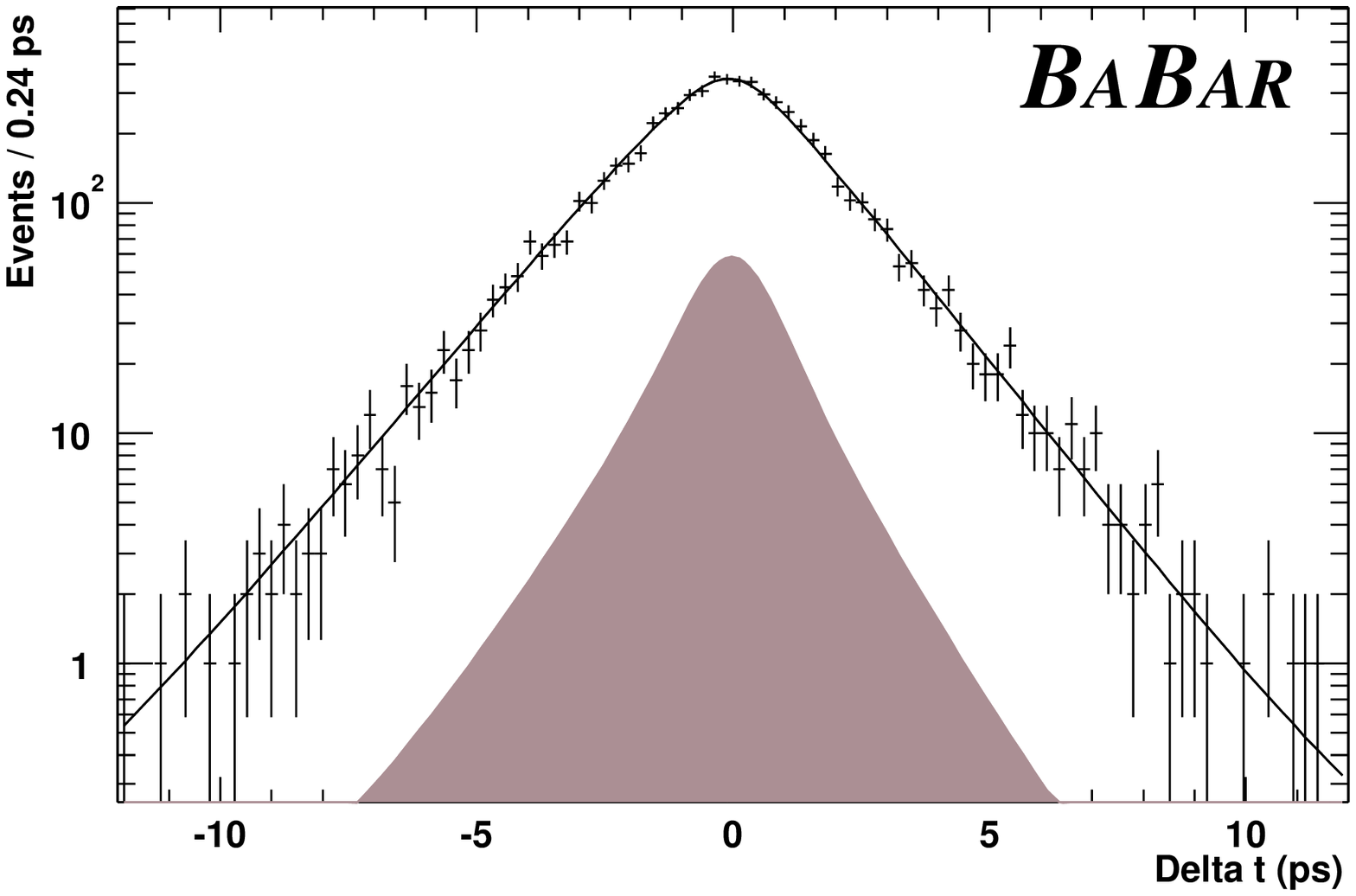,width=7cm}   & 
  \epsfig{file=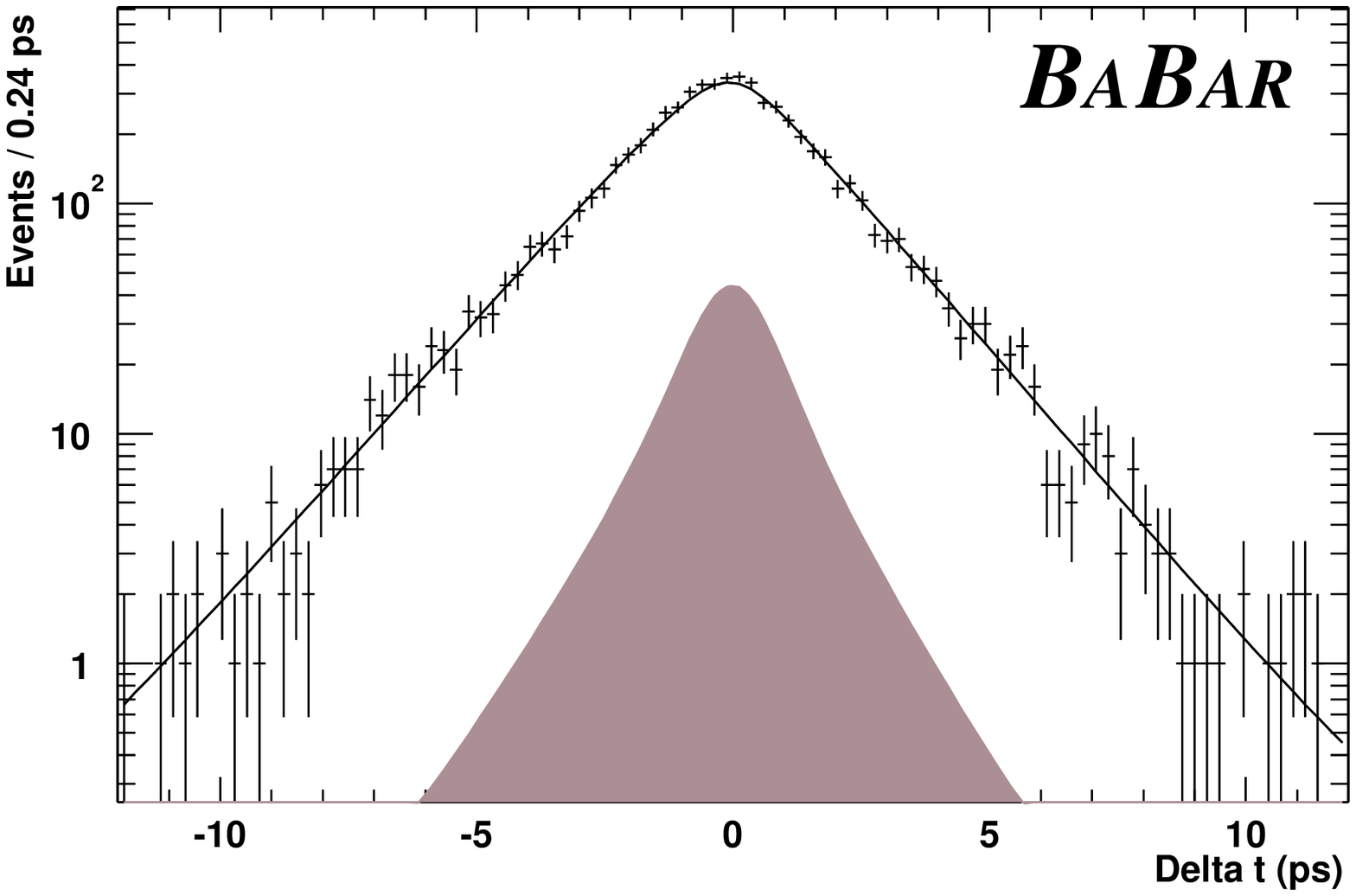,width=7cm}  \\
\end{tabular}
\caption{\it $\Delta t$ distributions for \Bz/\Bzb\ (left) and \Bu/\Bub\
  (right) candidates in the signal region (\mes$>5.27$\gevcc). The result of the lifetime
fit is superimposed. The background is shown by the hatched area.}
\label{fig:lifetime}
\end{center}
\end{figure}

\subsection{Time--dependent \BzBzb\ mixing}

A time-dependent \BzBzb\ mixing measurement 
requires the determination of the flavor of both \B\ mesons.
Considering the \BzBzb\ system as a whole, one can classify the tagged events 
as {\em mixed} or {\em unmixed} depending on whether the $B_{tag}$ is tagged 
with the same flavor as the $B_{rec}$ or with the opposite flavor. 
\par
From the time-dependent rate of mixed ($N_{mix}$) and unmixed
($N_{unmix}$) events, the mixing asymmetry 
$a(\Delta t) = (N_{unmix}-N_{mix})/(N_{unmix}+N_{mix})$ is
calculated as a function of $\Delta t$ and fit to the expected
cosine distribution.
A likelihood fit with 34 free parameters is performed.
The free parameters are the same as in the \stwob\ fit, apart from \stwob\ itself and the 
background to the \CP\ eigenstates which are not used, and \deltamd, which is, of course, floated.

\begin{figure}[htb]
\begin{center}
\begin{tabular}{lr}   
  \epsfig{file=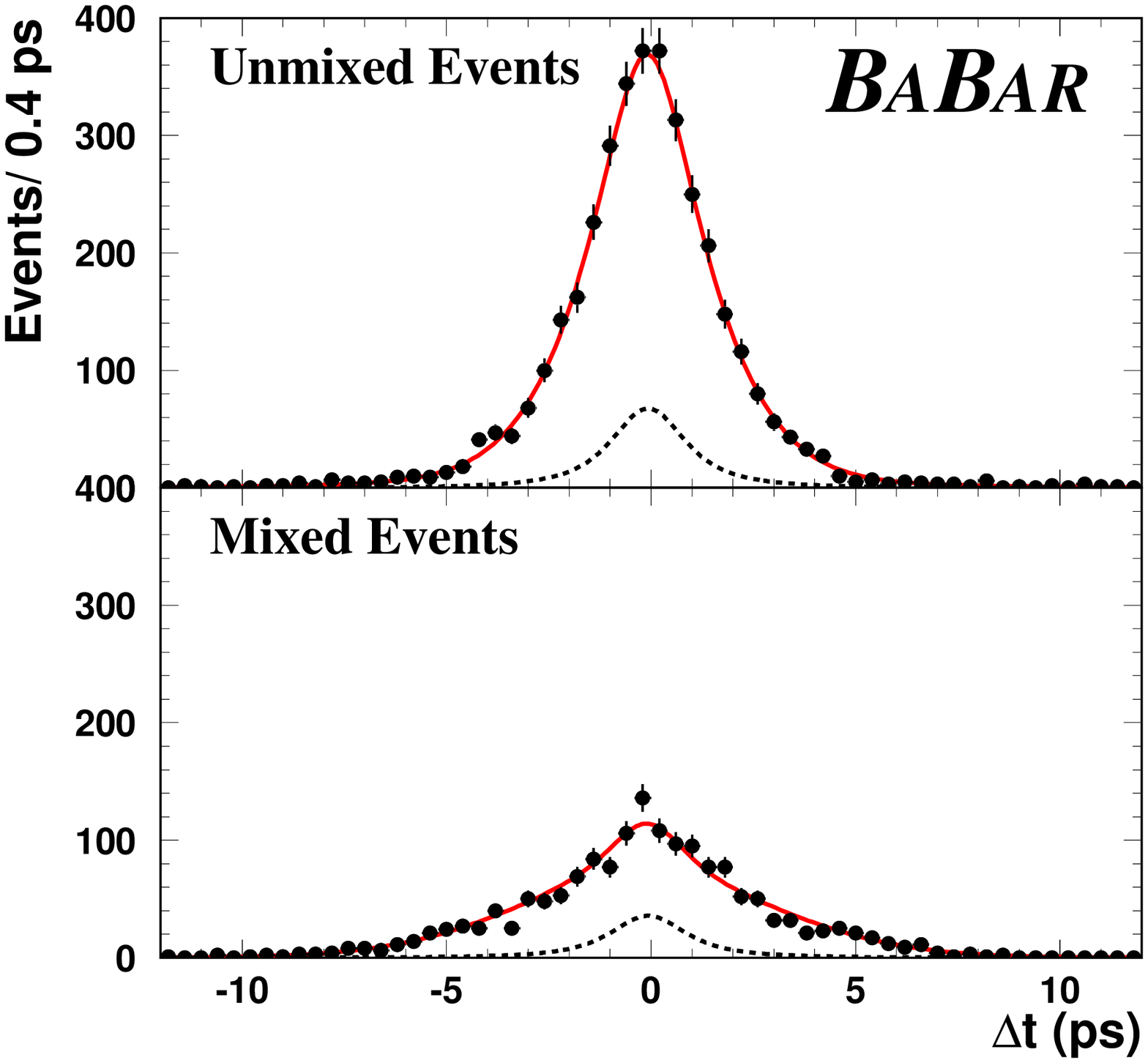,width=7cm}  &
  \epsfig{file=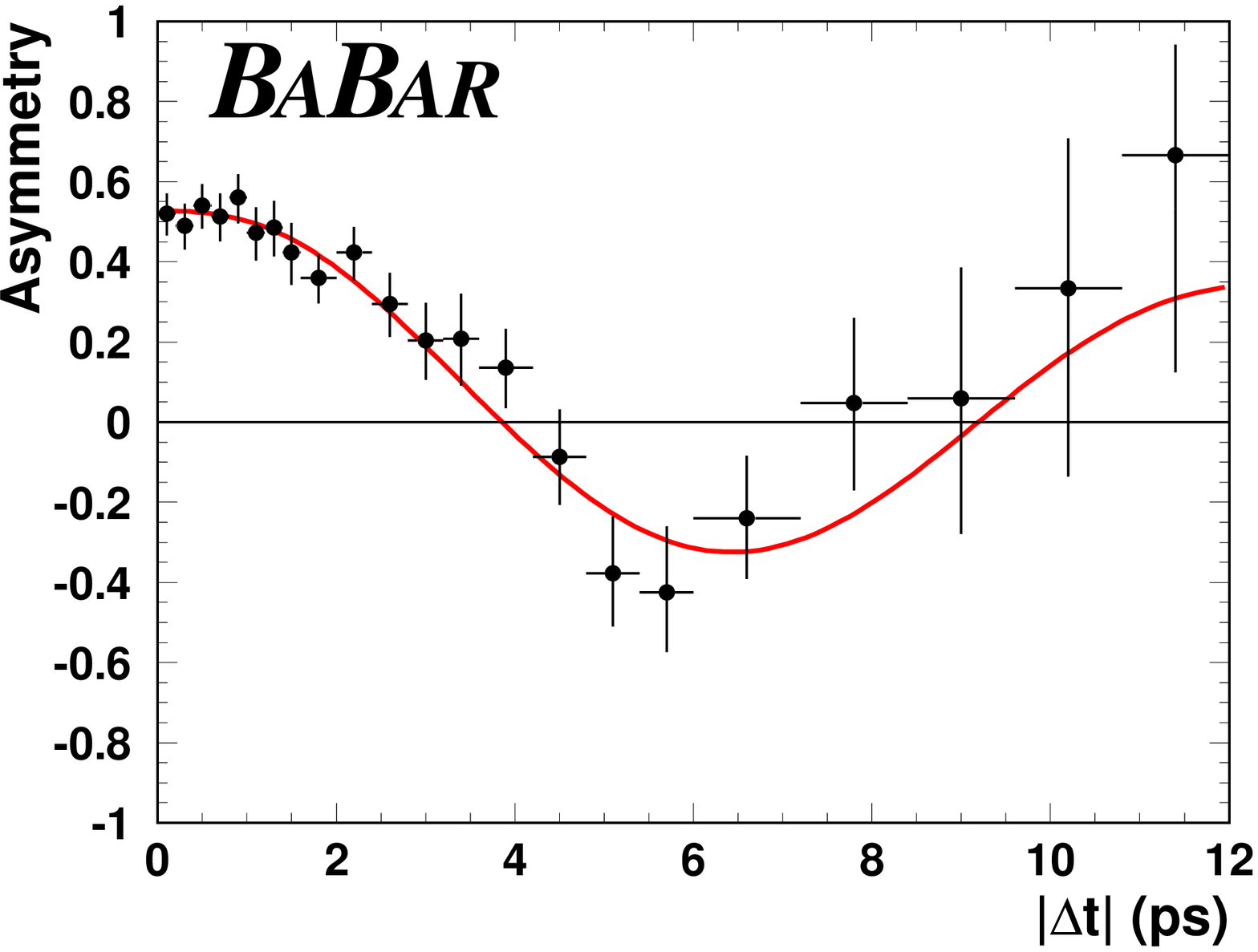,   width=8.4cm}  \\
\end{tabular}
\caption{\it \deltat\ distribution for mixed and unmixed events (left) and time-dependent asymmetry $a(\Delta t)$ between unmixed and
      mixed events (rigth)}
\label{fig:mixing}
\end{center}
\end{figure}
Fig. \ref{fig:mixing}
shows the \deltat\ and $a(\Delta t) = (N_{unmix}-N_{mix})/(N_{unmix}+N_{mix})$ distributions with the fit result superimposed. 

\subsection{Results}

The preliminary results for the \B\ meson lifetimes are
$\tau_{\Bz} = 1.546 \pm 0.032\ {\rm (stat)} \pm 0.022\ {\rm (syst)}\ \ps$ and
$\tau_{\Bu} = 1.673 \pm 0.032\ {\rm (stat)} \pm 0.022\ {\rm (syst)}\ \ps$
and for their ratio is
$\tau_{\Bu}/\tau_{\Bz} = 1.082 \pm 0.026\ {\rm (stat)}\ \pm 0.011\ {\rm (syst)}.$
\par
We measure the \BzBzb\ oscillation frequency:
$  \Delta m_d  =  0.519 \pm 0.020\ ({\rm stat})  \pm 0.016  ({\rm
  syst})\  \hbar {\rm ps}^{-1}$

The above results are consistent with previous
measurements~\cite{PDG2000} and  are of
similar precision. The mixing result is compatible with 
a \babar\ measurement using di--leptons \cite{BabarPub0010}. 

\section{Conclusions}
We have presented \babar's first measurement of the \CP-violating 
asymmetry parameter \stwob\ in the $B$ meson system:
\begin{equation}
\result .
\end{equation}  
Our measurement is consistent with the world average $\stwob = 0.9\pm0.4$~\cite{PDG2000}, 
and is currently limited by the size of the \CP\ sample.
\par We have also presented time--dependent mixing and lifetime measurements, 
performed for the first time at the \FourS.

\end{document}